\documentclass{article}
\usepackage{spconf,amsmath,graphicx}
\usepackage{color}

\title{Multi-task self-supervised pre-training for music classification}
\name{\begin{tabular}{c}Ho-Hsiang Wu$^{1, \star}$, Chieh-Chi Kao$^{2}$, Qingming Tang$^{2}$, Ming Sun$^{2}$ \\ Brian McFee$^{1}$, Juan Pablo Bello$^{1}$, Chao Wang$^{2}$\end{tabular}
\thanks{$^{\star}$ Work done at Amazon}
\thanks{This work is partially supported by the National Science Foundation award \#1544753}}
  
\address{$^{1}$ Music and Audio Research Laboratory, New York University, USA \\
         $^{2}$ Alexa Speech, Amazon}
\begin{document}
\ninept
\maketitle
\begin{abstract}

Deep learning is very data hungry, and supervised learning especially requires massive labeled data to work well. Machine listening research often suffers from limited labeled data problem, as human annotations are costly to acquire, and annotations for audio are time consuming and less intuitive. Besides, models learned from labeled dataset often embed biases specific to that particular dataset. Therefore, unsupervised learning techniques become popular approaches in solving machine listening problems. Particularly, a self-supervised learning technique utilizing reconstructions of multiple hand-crafted audio features has shown promising results when it is applied to speech domain such as emotion recognition and automatic speech recognition (ASR). In this paper, we apply self-supervised and multi-task learning methods for pre-training music encoders, and explore various design choices including encoder architectures, weighting mechanisms to combine losses from multiple tasks, and worker selections of pretext tasks. We investigate how these design choices interact with various downstream music classification tasks. We find that using various music specific workers altogether with weighting mechanisms to balance the losses during pre-training helps improve and generalize to the downstream tasks.

\end{abstract}
\section{Introduction}
\label{sec:introduction}

Deep learning has shown great successes with end-to-end learned representations replacing hand-crafted features in various machine perception fields, including computer vision, natural language processing and machine listening, especially in supervised learning paradigm. However, unlike ImageNet for computer vision, which contains millions of labeled images, human annotated datasets for machine listening are usually small \cite{choi2016automatic}. Therefore, learning from limited labeled data \cite{kim2020one} is especially important. There are existing methods such as transfer learning \cite{choi2017transfer} and domain adaptation, where models learned from different tasks with larger datasets are transferred and fine-tuned to another task/domain, and unsupervised learning \cite{wulfing2012unsupervised, schneider2019wav2vec, baevski2019vq}, such as generative models \cite{oord2016wavenet, kumar2019melgan}, where data distribution is often learned through reconstruction of the signal.

Self-supervised learning \cite{cramer2019look, chen2020simple, chen2020big, gfeller2020spice}, as one sub-field of unsupervised learning, exploits the structure of the input data to provide supervision signals. It has become more popular in recent years, showing good improvement in multiple fields. For self-supervised learning, raw signals are transformed, and models are optimized with reconstruction or contrastive losses against original signals, where preserving of temporal or spatial data consistency is assumed for learning meaningful representations. These representations are proven useful to generalize and solve downstream tasks. On the other hand, multi-task learning \cite{hung2019multitask} improves generality by solving multiple tasks altogether during training, while weighting mechanisms among the losses from each task are crucial \cite{kendall2018multi, gong2019comparison}. Self-supervised and multi-task learning techniques are combined and applied to the speech domain, and they have shown success in \cite{pascual2019learning, ravanelli2020multi}, where reconstruction of various hand-crafted features are used for pre-training, and further learned representations are evaluated with downstream emotion recognition and automatic speech recognition (ASR) tasks.


Similar to speech, music is also a highly structured audio signal. There are many hand-crafted features designed specifically for music to solve various music information retrieval (MIR) tasks. In this paper, we are interested in applying self-supervised and multi-task learning methods for pre-training music encoders. We explore various design choices including encoder architectures, weighting mechanisms to combine losses from pretext tasks, and worker selections to reconstruct various music specific hand-crafted features, such as Mel-frequency cepstral coefficients (MFCCs) for timbre \cite{de2012enhancing}, Chroma for harmonic \cite{ellis2007classifying}, and Tempogram \cite{grosche2010cyclic} for rhythmic attributes. Our main contributions are 1. provide suggestions on best design choice among all the variations from our experiments, and 2. investigate how different selections of pretext tasks interact with the performance of downstream music classification tasks, including instrument, rhythm and genre.



\section{Method}
\label{sec:method}

\begin{figure}[ht]
\centering
\includegraphics[width=\linewidth]{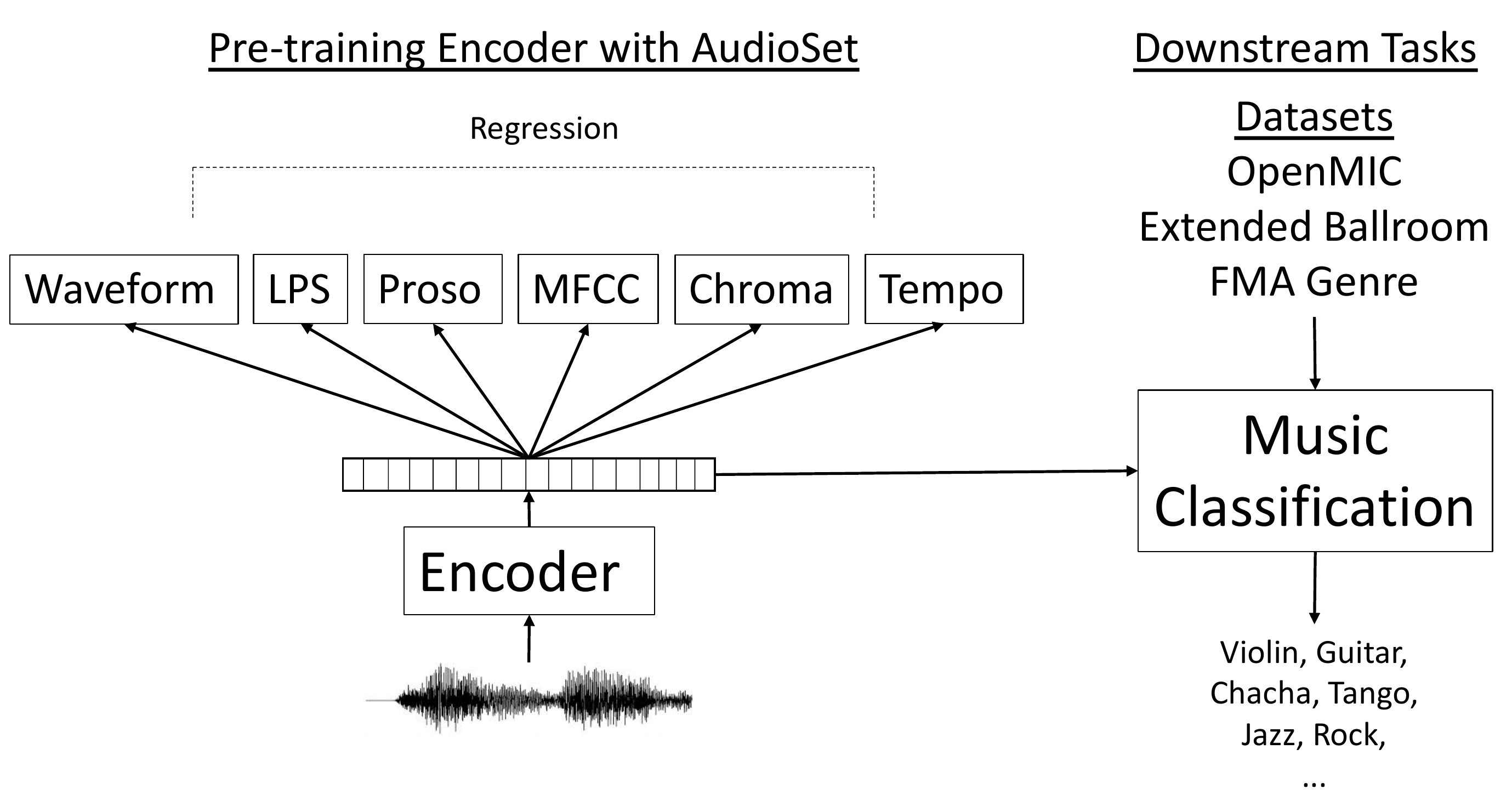}
\caption{Diagram of multi-task self-supervised encoder pre-training and downstream music classification evaluation.}
\label{fig:diagram}
\end{figure}

A two-stage approach involving unsupervised or self-supervised pre-training and supervised learning for training to evaluate on downstream tasks is commonly adopted \cite{cramer2019look, chen2020simple, pascual2019learning, ravanelli2020multi} in recent literature, especially in the context of limited labeled data, where representation learning is key. In order to evaluate the effectiveness of the pre-training, simple linear or multi-layer perceptron (MLP) classifiers are usually used where the pre-trained encoders are required to capture meaningful representations to perform well on linear separation evaluation tasks.


\subsection{Multi-task self-supervised pre-training}
As shown in Figure \ref{fig:diagram}, we combine self-supervised and multi-task learning ideas for pre-training. Raw audio inputs are passed through multiple encoding layers, and outputs are two dimensional representations with temporal information. These encoded representations are then used for solving pretext tasks via workers including waveform reconstruction, and prediction of various popular hand-crafted features used in MIR to guide the learning jointly. 


\subsection{Downstream task training scenarios}
\label{subsec:scenarios}
After pre-training, we remove the workers, and feed the encoder outputs to MLP classifiers for downstream tasks. We adopt three training scenarios proposed in \cite{pascual2019learning}: 1. \textbf{Supervised}: Initialize the encoder weights randomly and train from scratch on the downstream datasets directly. 2. \textbf{Frozen}: Treat the pre-trained encoder as feature extractor with frozen weights, concatenate the feature extractor with trainable MLP classifiers and only optimize the classifier weights. 3. \textbf{Fine-tuned}: Initialize the encoder with pre-trained weights and fine-tune the encoder with downstream tasks altogether.

\section{Experimental Design}
\label{sec:experimental_design}

We experiment with various design choices during pre-training including 1. Encoder architectures, 2. Pretext tasks for worker selections, 3. Weighting mechanisms for losses from pretext tasks. We provide more details on the downstream evaluations and data usage for both pre-training and downstream tasks in section \ref{subsec:downstream_eval} and \ref{subsec:data}.

\subsection{Encoder architectures}
We compare two encoder architectures proposed in two relevant studies in speech domain which inspire our work. We refer the two encoder architectures as PASE \cite{pascual2019learning} and PASE+ \cite{ravanelli2020multi}, respectively.

1. \textbf{PASE}: We use the same encoder architecture as the original PASE work \cite{pascual2019learning} with source code implementation\footnote{https://github.com/santi-pdp/pase}. The first layer is based on SincNet \cite{ravanelli2018speaker}, where the raw input waveform is convolved with a set of parameterized Sinc functions implementing rectangular band-pass filters. The authors claim that SincNet has fewer parameters and provides better interpretability. SincNet layer is followed by 7 one-dimensional convolutional blocks, batch normalization \cite{ioffe2015batch}, and multi-parametric rectified linear unit activation \cite{he2015delving}. We use the same model parameters as provided in the original work including kernel widths, number of filters, and strides. The set of parameters for convolutional layers emulates a 10ms sliding window. 

2. \textbf{PASE+}: PASE+ \cite{ravanelli2020multi} improves upon PASE \cite{pascual2019learning} by adding skip connections and Quasi-Recurrent Neural Network (QRNN) \cite{bradbury2016quasi} layers to capture longer-term contextual information. QRNN layers consist of interleaved convolutional layers with RNN layers to speed up training with parallel optimization, while maintaining compatible performance.

\subsection{Pretext tasks worker selections}
Inspired by the original PASE \cite{pascual2019learning} work, we select waveform reconstruction, log power spectrum (LPS) and prosody features as baseline workers. We then choose three popular hand-crafted features in MIR field including MFCC, Chroma, and Tempogram as mixed-in workers. For waveform reconstruction, encoder layers are applied in reverse order to decode embeddings and optimized with mean absolute error (MAE) loss. For all the other workers, we use MLP with convolutional layers, and mean squared error (MSE) loss.



Waveform, LPS, and MFCC are commonly used in machine listening. Chroma is inspired from western 12-tone theory which frequencies are folded into 12 bins as one octave. Tempogram \cite{grosche2010cyclic} takes local auto-correlation of the onset strength envelope. As used in \cite{pascual2019learning}, prosody features include zero crossing rate (ZCR), energy, voice/unvoice probability and fundamental frequency (F0) estimation, resulting in 4 features concatenated along with temporal dimension. For LPS, MFCC, Chroma, Tempogram and prosody, we use librosa\footnote{https://github.com/librosa/librosa} implementations with hop\_length = 160, n\_fft = 2048, sr = 16000 in order to align each hop as 10ms to match encoder parameters, with other default parameters.



\subsection{Weighting mechanisms}
We explore two weighting mechanisms to combine losses from each worker during pre-training. 1. \textbf{Equal weighted} by simply sum up losses from different workers for backpropagation. 2. \textbf{Re-weighted} by taking the validation losses per worker of the first 10 epochs from equal weighted training, averaging the loss per worker, taking the reciprocal as the new weights and applying those to retrain from scratch. The intuition is that the losses from each worker will then contribute more equally during backpropagation optimization.


\subsection{Downstream evaluation}
\label{subsec:downstream_eval}
After pre-training, we remove the workers for pretext tasks and concatenate the output of the encoder with a simple MLP classifier. The input layer of the MLP is to take mean pooling across temporal dimension, resulting in one 512 dimension embedding, followed by 1 fully connected layer to adapt to output dimensions corresponding to the number of classes of each downstream dataset. We train with three scenarios discussed in section \ref{subsec:scenarios}, including supervised, frozen and fine-tuned, all with the same hyper-parameters, Adam optimizer \cite{kingma2014adam} with initial learning rate as 0.001 and early stopping criteria with patience value of 10 on validation loss. We run 10 trials for each experiment in this paper to get statistically meaningful results.

\subsection{Data}
\label{subsec:data}

\subsubsection{AudioSet for pre-training}
We use clips in AudioSet \cite{gemmeke2017audio} with "Music" label for pre-training. We are able to acquire \~{}2M (97\% of the original AudioSet data) clips, within which there are \~{}980k clips labeled with "Music". We randomly select 100k for pre-training, resulting in \~{}83 hours of data.

\subsubsection{Datasets for downstream evaluation}
OpenMIC \cite{humphrey2018openmic}, Extended Ballroom \cite{marchand2016extended} and FMA Small (FMA) \cite{defferrard2017fma}, three publicly available classification datasets are used for downstream evaluation as representative samples of well-known MIR tasks. These datasets range from different number of clips, clip duration, and number of classes. For all three datasets, we report macro F1 scores as shown in the figures. 


\begin{enumerate}
  \item OpenMIC \cite{humphrey2018openmic}: OpenMIC is a multi-label instrument classification dataset containing 15k samples total with provided train/valid/test splits as well as masks for strong positive and negative examples for each class. We follow similar setup as the official baseline\footnote{https://github.com/cosmir/openmic-2018} by training 20 binary classifiers.
  \item Extended Ballroom \cite{marchand2016extended}: Extended Ballroom (4k samples) is a multi-class dance genre classification dataset. We follow the same setup as \cite{jeong2017dlr} by removing 4 categories due to dataset imbalance, resulting in only using 9 categories.
  \item FMA Small \cite{defferrard2017fma}: FMA Small (8k samples) is a multi-class music genre classification dataset with 8 genre categories.
\end{enumerate}

\section{Results and discussions}
\label{sec:results}

We first show results of encoder choices and whether pre-training helps. All workers (waveform (W), LPS (L), prosody (P), MFCC (M), Chroma (C) and Tempogram (T), where WLP are also referred to as baseline) and frozen scenario are used. We then dive deeper into the effects of different weighting mechanisms, and ablation study of worker selections, for which we also report results in frozen scenario. Finally, we investigate whether fine-tuning further improves performance.

\subsection{Encoder architectures}
\label{subsec:encoder_choices}

\begin{figure}[ht]
\centering
\includegraphics[width=\linewidth]{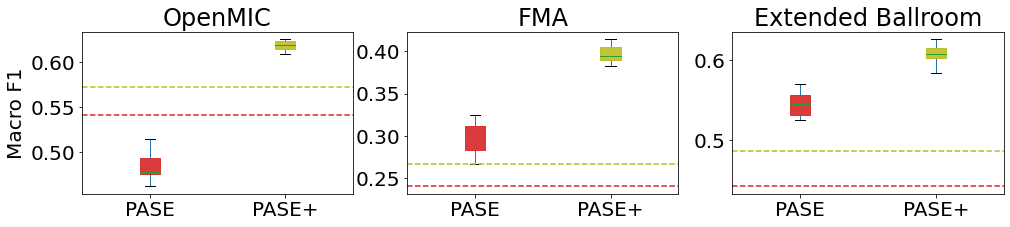}
\caption{Comparisons of encoder architectures (PASE vs PASE+). Left, center, and right figures are Macro F1 metrics on different downstream tasks of frozen scenario. Red and green dotted lines represent PASE and PASE+ encoder with supervised training (scenario 1) directly on downstream dataset from scratch.}
\label{fig:info_max}
\end{figure}

From Figure \ref{fig:info_max}, we observe that for all three downstream tasks, PASE+ outperforms PASE. This is not surprising as PASE+ is a more powerful encoder with \~{}8M parameters, skip-connection and QRNN layer, and PASE has only \~{}6M parameters and basic convolutional layers. This confirms with the findings from original PASE+ \cite{ravanelli2020multi} work applied to speech data.

The dotted lines are trained supervisedly (scenario 1) from scratch directly on the downstream tasks with random weights initialization. It shows that pre-training in general helps to initialize the encoder weights better, resulting in better performance on downstream tasks. One exception is PASE for OpenMIC, we hypothesize that it is because OpenMIC already contains enough data to train PASE encoder (with less capacity) from scratch well, which is not the case for PASE+. This shows that pre-training for encoders with larger capacities is especially helpful when evaluating on downstream tasks with limited labeled data. We conducted experiments using PASE+ through out the remaining paper as it's a better encoder for our tasks.



\subsection{Weighting mechanisms}

\begin{figure}[!ht]
\centering
\includegraphics[width=\linewidth]{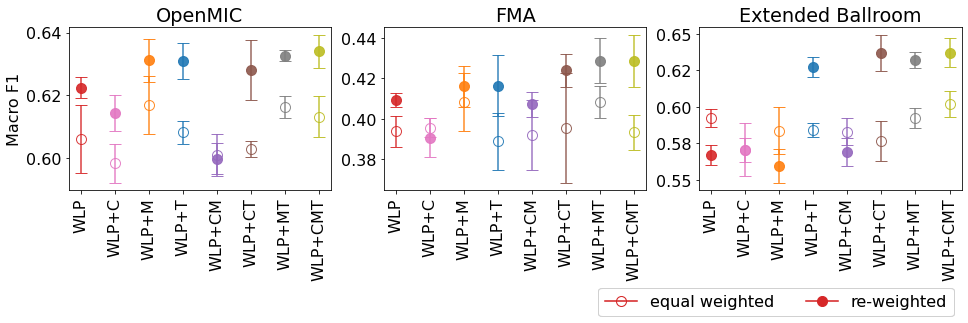}
\caption{Comparisons of equal weighted vs re-weighted for different worker selections on all downstream tasks. PASE+ encoder architecture is used with frozen scenarios. Y-axis is Macro F1 classification metrics. X-axis are labeled with WLP (waveform, LPS, and prosody), M (MFCC), C (Chroma), and T (Tempogram). No filled and filled color represent equal weighted and re-weighted mechanisms correspondingly. From all trials, circles represent mean while the length of the bar represents standard deviation.}
\label{fig:weighted}
\end{figure}

In Figure \ref{fig:weighted}, we show results comparing equal weighted and re-weighted mechanisms with different worker selections during pre-training. We see that re-weighted mechanism (filled color) helps to boost the influences from various workers to the performance of downstream tasks in general. For Extended Ballroom on the right especially, we see clearly that results with workers containing Tempogram are improved by a large margin.

\begin{figure}[ht]
\centering
\includegraphics[width=\linewidth]{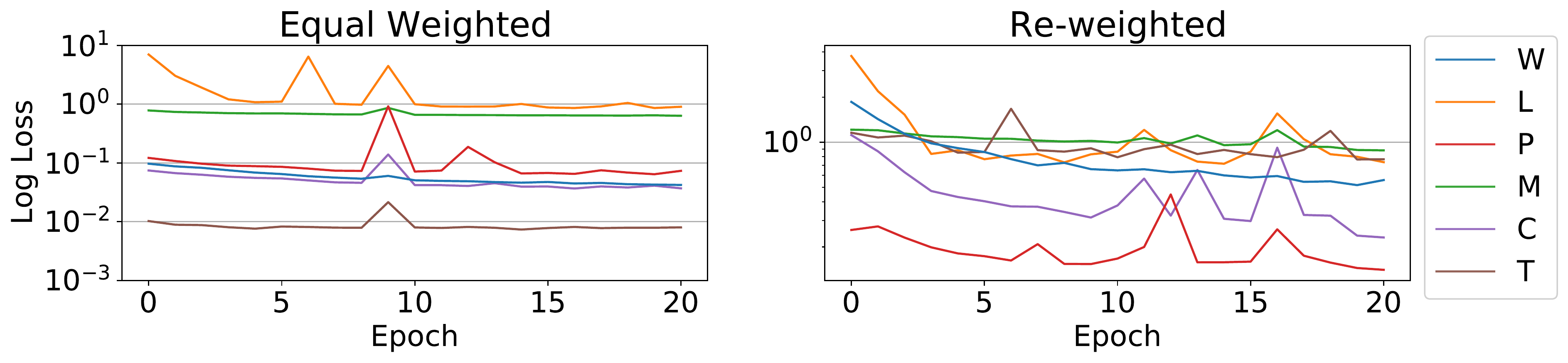}
\caption{Log loss per worker for first 20 epochs. X-axis is number of epochs. On the left is equal weighted. On the right is re-weighted where loss weights are balanced using reciprocal of mean losses per worker from equal weighted pre-training.}
\label{fig:loss_weight}
\end{figure}

We further examine losses per worker during pre-training as shown in Figure \ref{fig:loss_weight}. We can see that with equal weighted on the left, LPS (L) almost dominates all losses and Tempogram (T) worker loss contributes the least with two orders of magnitude smaller, but for re-weighted on the right, each worker contributes more equally.

\subsection{Pretext tasks worker selections}
\begin{figure}[ht]
\centering
\includegraphics[width=\linewidth]{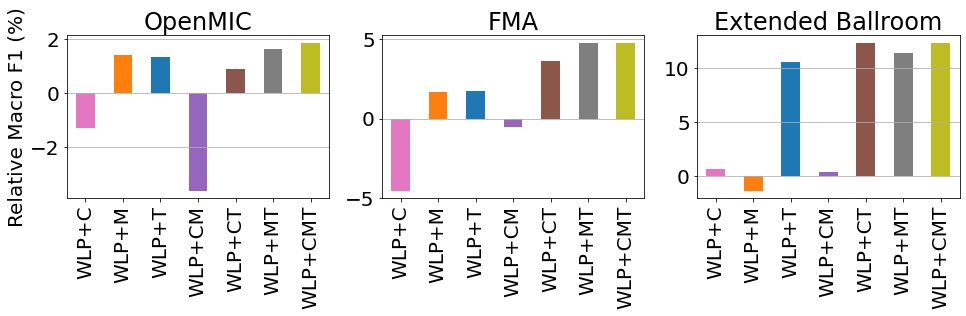}
\caption{Relative improvement (\%) of different additional music specific workers included during pre-training compared to WLP on different downstream tasks.}
\label{fig:ablation}
\end{figure}

Figure \ref{fig:ablation} shows the relative difference in accuracy by including different workers over the WLP baseline. We observe that different worker selections affect variously to different downstream tasks. Tempogram helps the most across all different combinations especially for Extended Ballroom. MFCC is usually important for most of the downstream tasks as it captures the low-level attributes differentiating instrument and genre. Chroma is however at a disadvantage, especially for OpenMIC, since Chroma is designed to normalize for timbre, which is important for instrumentation. MFCC only hurts slightly on Extended Ballroom as it brings together different dance genres with similar timbre, and separates music from same dance genre that changes in timbre.

These variations can be further compensated to show improvement across all tasks by using all workers as shown on the right most of each subplot in Figure \ref{fig:ablation}. We observe relative improvement adding all workers compared to WLP baseline by 1.9\%, 4.5\% and 14\% on OpenMIC, FMA and Extended Ballroom datasets respectively. This indicates that workers complement each other, and the encoders are able to use signals from diversified workers to generalize better to various downstream tasks.

\begin{figure}[ht]
\centering
\includegraphics[width=\linewidth]{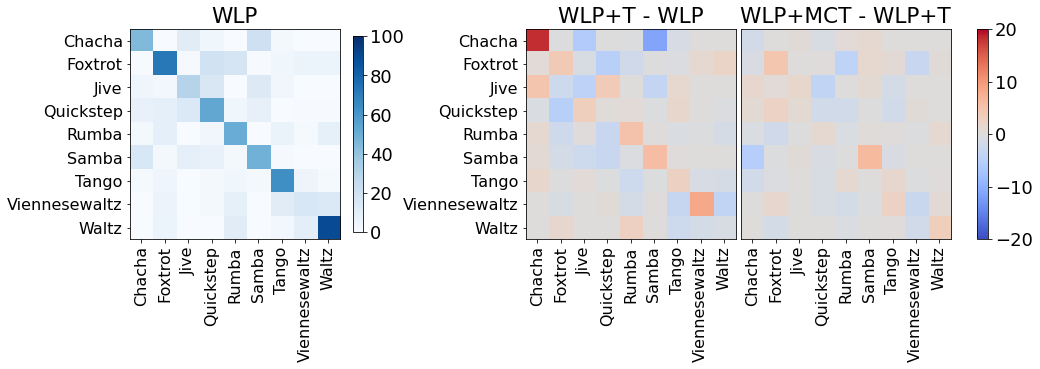}
\caption{Confusion matrices of Extended Ballroom. On the left is WLP baseline. On the right are the differences between WLP+T and WLP, and WLP+MCT and WLP+T. Red and blue colors indicate positive and negative changes respectively.}
\label{fig:cm_extended_ballroom}
\end{figure}

\begin{figure}[ht]
\centering
\includegraphics[width=\linewidth]{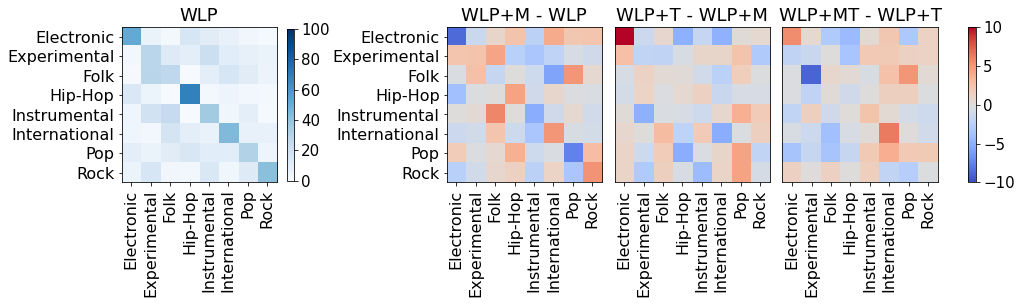}
\caption{Confusion matrices of FMA. On the left is WLP baseline. On the right are the differences between WLP+M and WLP, WLP+T and WLP+M, and WLP+MT and WLP+T. Red and blue colors indicate positive and negative changes respectively.}
\label{fig:cm_fma}
\end{figure}

We then show confusion matrices of Extended Ballroom and FMA in Figure \ref{fig:cm_extended_ballroom} and \ref{fig:cm_fma}. In Figure \ref{fig:cm_extended_ballroom}, we show the difference between WLP + T and WLP, and observe that adding Tempogram helps differentiate Chacha with Jive and Samba, which differ in rhythm and tempo, as well as Foxtrot with Quickstep, and Viennesewaltz with Waltz, as the two pairs of dance genres originate from similar music playing in different speed. Adding MFCC and Chroma further helps differentiate Foxtrot with Rumba and Viennesewaltz as additional timbre cues are provided.

In Figure \ref{fig:cm_fma}, we observe that even adding MFCC (WLP+M - WLP) helps in general as hypothesized, however, it misclassifies Electronic with Hip-Hop and International, and Pop with Hip-Hop and Rock, as there might be similar instruments used in these genres, resulting in similar timbre. Adding Tempogram (WLP+T - WLP+M) corrects the mistakes made on Electronic and Pop genres, but misclassifying International with Folk and Instrumental. Finally, adding both workers (WLP+MT - WLP+T) provides further improvements upon MFCC and Tempogram only. In general we observe improvements with positive values (red) in diagonal and negative (blue) in off-diagonal.



\subsection{Frozen versus fine-tuned}

\begin{figure}[ht]
\centering
\includegraphics[width=\linewidth]{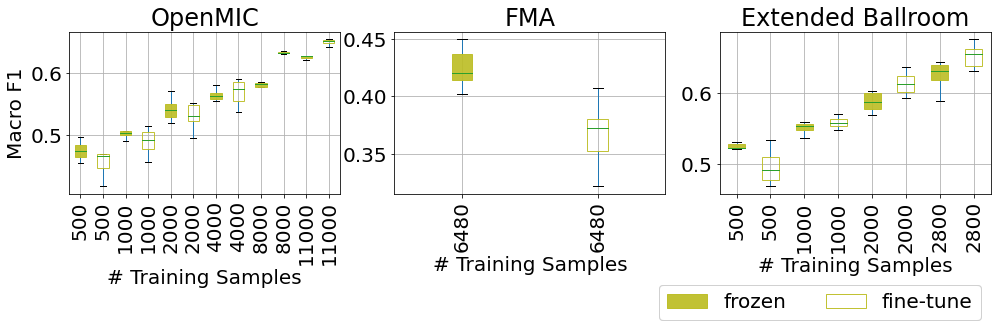}
\caption{Comparisons of frozen and fine-tuned on \# of training samples for different downstream tasks.}
\label{fig:train_num}
\end{figure}

In Figure \ref{fig:train_num}, we plot frozen (filled) versus fine-tuned (no filled) with re-weighted mechanisms and all workers used during pre-training. By using all available training examples, both Extended Ballroom (2.8k) and OpenMIC (11k) show further improvement with fine-tuning, while FMA does not. We hypothesize that this is because each downstream task requires different number of samples for fine-tuned to work well. For FMA, we just do not have enough training samples. We further reduce number of samples used for training OpenMIC and Extended Ballroom as shown in Figure \ref{fig:train_num}, where we see clear reverting behavior around 8k (OpenMIC) and 1k (Extended Ballroom) that fine-tuning stops to outperform frozen.




\section{Conclusion}
\label{sec:conclusion}

In this paper, we explore different design choices for pre-training music encoders with multi-task and self-supervised learning techniques, and show that this method, when combined with different encoder architectures, generally benefits for downstream tasks. The improvement is clearer and more stable when (\# unlabeled data / \# labeled data) is larger. We also show that each type of pretext task provides different and complementary information, re-weighted mechanism helps the encoder to better learn different cues provided from each task, and fine-tuning can further improve performance.

For future work, we are interested in applying this pre-training technique to various encoders, adding more audio specific features, and explore other unsupervised and self-supervised learning ideas such as wav2vec \cite{schneider2019wav2vec} as pretext tasks. We are also interested in including more diverse downstream tasks such as music tagging, and chord recognition (Chroma should be more effective in this task) for evaluation. We think that this pre-training technique can be applied to a large varieties of music encoders and generalize to different downstream music tasks, especially those with limited labeled data.


\vfill\pagebreak

\bibliographystyle{IEEEbib}
\bibliography{refs}

\end{document}